\documentclass[12pt]{iopart}

\usepackage{iopams} 
\usepackage{graphicx}
\usepackage{url}
\usepackage{multirow}
\usepackage{xcolor}
\usepackage{tabularray}

\renewcommand{\t}{\hat{\pmb t}}
\newcommand{\n}{\hat{\pmb \kappa}}
\renewcommand{\b}{\hat{\pmb \tau}}

\newcommand{\eightchar}{%
  \begingroup\normalfont
  \includegraphics[height=\fontcharht\font`\B]{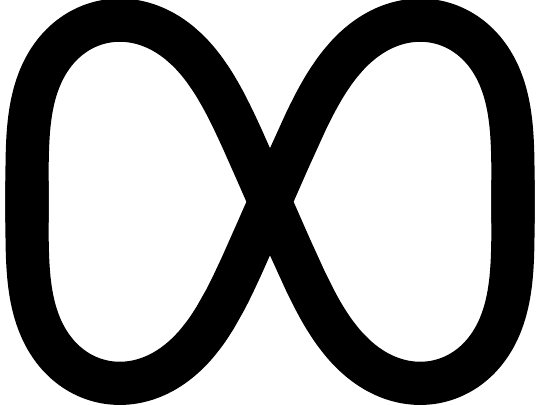}%
  \endgroup
}

\makeatletter
\newcommand{\iotaslash}{{\mathpalette\iotaslash@\relax}}

\newcommand{\iotaslash@}[2]{
  \raisebox{-0.55\height}[0pt][0pt]{%
    \rotatebox[origin=Bl]{20}{$\m@th#1{\mkern-1mu\mathchar'26}$}%
  }%
  \mkern-9.0mu\nonscript\mkern-1mu
  \iota
  \mkern1mu
}
\makeatother

\begin{document}

\title[Back to the Figure-8 Stellarator]{Back to the Figure-8 Stellarator}

\author{G. G. Plunk$^1$, M. Drevlak$^1$, E. Rodr\'{i}guez$^1$, R. Babin$^2$, A. Goodman$^1$ and F. Hindenlang$^2$}

\address{$^1$Max Planck Institute for Plasma Physics, Greifswald, Germany}
\address{$^2$Numerical Methods in Plasma Physics, Max Planck Institute for Plasma Physics, Garching, Germany}

\ead{ggplunk@ipp.mpg.de}
\vspace{10pt}

\begin{abstract}
The first stellarator design was a simple tube of plasma twisted and closed on itself in the form of a figure-8.  The line of such devices, however, was quickly ended over concerns related to plasma stability.  We revisit the figure-8 concept, re-imagined as a modern optimized stellarator, and find the potential for a high degree of stability, as well as exceptionally simple construction.  In particular, the design that we find admits planar coils, and is the first quasi-isodynamic stellarator design to have this property.  Our work is made possible by recent theoretical progress in the near-axis theory of quasi-isodynamic stellarators, combined with fundamental progress in the numerical solution of three-dimensional magnetohydrodynamic equilibria that cannot be well represented using traditional cylindrical coordinates.
\end{abstract}

%
%
%
%
%

\section{Introduction}

In 1951, Lyman Spitzer had an idea for a device to achieve controlled thermonuclear fusion.  The form he envisioned had the shape of a figure-8, an endless tube of plasma, bent around and closed onto itself.\footnote{According to folklore, the idea for the figure-8 may have occurred on Spitzer's ski trip to Aspen, perhaps even on a ski lift, but some authors state that the idea came later in Princeton \cite{bromberg, stix-early-stellarator}.}  Although this idea was realized in a number of experiments during the following years \cite{goldman1953modela,Model-C-Report-PMS-19}, it was decided that the figure-8 stellarator was unstable and that it would be better to switch over to stellarators that achieve rotation of magnetic field lines via helically wound coils \cite{spitzer1954problems}, {\em e.g.} so-called classical stellarators.  These, it was predicted, would be stabilized by magnetic shear, and were also expected to be easier to construct than the figure-8 because the plasma volume would have a simpler, more toroidal shape \cite{spitzer-1958, bishop-project-sherwood}.

Times have changed.  The modern stellarator, which can be thought of as a descendant of the classical stellarator, is carefully sculpted using computational optimization techniques, resulting in shapes that are impressive to behold but hard to build.  Furthermore, although stellarators can exhibit a high degree of stability, examples that are sufficiently well-optimized for plasma confinement tend to have fairly low magnetic shear, and owe their stability more to other properties such as the so-called ``magnetic well''.  Thus, the justifications for turning away from the figure-8 do not seem to apply anymore.

It might seem improbable that a design discarded long ago could possibly stand up against modern stellarators.  Indeed, if the figure-8 were really a credible design, meeting the various requirements of a fusion device, why wouldn't it naturally arise from stellarator optimization?  A possible answer is that the figure-8, despite its relative simplicity, presents a technical challenge to investigate because its particular shape is not easily represented by the coordinates (cylindrical) that are at the heart of the equilibrium codes used for optimization.  It seems reasonable to assume that such codes use these coordinates because they suited the kind of stellarators that were of interest at the time when the codes were being developed, and the resounding success of stellarator optimization left no compelling reason to doubt this choice.

However, methods developed in recent years to construct new stellarator designs, mostly based on near-axis theory, have attracted the attention of researchers to a broader range of stellarator shapes.  Additionally, developers of recent stellarator equilibrium solvers \cite{gvec-2019, dudt_DESC_2020}, aware of the potential fundamental limitations of arbitrarily fixed coordinates, built in capabilities to explore alternative choices.  The time is right, we think, to revisit the figure-8.

In this paper, we make a first investigation of Spitzer's figure-8, re-imagined as a modern quasi-isodynamic (QI) stellarator \cite{gori-lotz-nuehrenberg, Helander_2009,nuerenberg-ppcf-2010}.  This is enabled by recent theoretical progress in the near-axis theory of QI stellarators \cite{plunk_landreman_helander_2019, camacho-mata_plunk_2022, plunk2024-QI,rodriguez-plunk-second-order-QI-construction-2024}, as well as the extension of the three dimensional magnetohydrodynamic equilibrium solver GVEC to use flexible user-specified coordinates to handle virtually any shape around a closed space curve \cite{Hindenlang_GVEC-Frenet_2022}.  The present work therefore serves as a first demonstration of the potential for optimization of `exotic' stellarator configurations, for which progress has so far been difficult or impossible.  More extensive work will become possible as GVEC is integrated with stellarator optimization tools \cite{Babin-py-gvec-2024} and as other numerical tools are adapted to interface with it.

The paper is organized as follows.  In section \ref{sec:nae_construction} we describe the near-axis method for constructing approximately QI configurations of the figure-8 type.  This includes the shape of the magnetic axis, and the elliptical shaping of the magnetic surfaces.  In section \ref{sec:figure-eight-sequence}, we describe what makes the figure-8 special, key geometric measures and how they affect the properties of the first-order near-axis solutions.  Section \ref{sec:well-opt} describes the numerical optimization of a set of configurations for a vacuum magnetic well, to demonstrate the tendencies of the figure-8 to support magnetohydrodynamic stability.  This section includes a discussion of the underlying geometric reasons for these tendencies, from the perspective of near-axis theory.  In section \ref{sec:QI-figure-eight-planar-coils}, we turn attention to a single configuration to demonstrate compatibility with a simple planar coil set, after which we conclude with a general discussion.

\section{Near-axis construction of a QI Figure-8 stellarator} \label{sec:nae_construction}

Low-field-period ($N$) configurations, including those $N=1$ and $N=2$ examples with a figure-8 appearance, have arisen using the theory of the near-axis expansion (NAE) \cite{plunk_landreman_helander_2019, camacho-mata_plunk_2022, Jorge_2022}.  Such solutions can be ``directly constructed'' \cite{landreman-sengupta,plunk_landreman_helander_2019} by evaluating the asymptotic description near the magnetic axis, at a specified distance (average minor radius).  The resulting plasma boundary shape can be provided as input to a magnetohydrodynamic (MHD) equilibrium code like GVEC for further investigation.

Based on the methods described in \cite{plunk_landreman_helander_2019, plunk2024-QI}, we will perform this construction initially at first order in the expansion; to understand the issue of stability we will later make an analysis at second order (see Section \ref{eduardo-sec}).  

\subsection{Magnetic axis shape}

The NAE for QI stellarators differs significantly from that for quasi-symmetric (QS) stellarators \cite{boozer1983,nuehrenberg-zille,rodriguez2020necessary}. For QI, controlling the geometry of the magnetic axis curve to produce flattening points (points of zero curvature) is a significant first hurdle.  We choose to obtain the magnetic axis curve by solving the Frenet-Serret equations \cite{frenet1852courbes, animov2001differential} providing the curvature, $\kappa$, and torsion, $\tau$, as functions of the arc length $\ell$ along the curve,

\begin{equation}
    \frac{d\mathbf{x}_0}{d\ell}=\t,\quad
    \frac{d\hat{\pmb t}}{d\ell}=\kappa\n,\quad
    \frac{d\n}{d\ell}=-\kappa\t + \tau \b,\quad
    \frac{d\b}{d\ell}=-\tau\n,  \label{eqn:FS_eqns}       
\end{equation}
where ${\bf x}_0$ is the axis curve, and $\hat{\pmb t}$ is its tangent.  Note that twice-differentiability of the curve is assured, and further smoothness is inherited from smooth choices of inputs $\kappa$ and $\tau$.  However, not all functions $\kappa$ and $\tau$ yield a closed curve, so optimization is required \cite{garren-boozer-1}.  The method used here has been described elsewhere \cite{plunk-simons-2023} and further technical details will be provided in a separate publication \cite{plunk2024-QI}.

Because of the points of zero curvature that lie along the axis curves, the frame $(\t, \n, \b)$ is not actually the conventionally defined Frenet frame, but the {\em signed} Frenet frame \cite{plunk_landreman_helander_2019,rodriguez-plunk-second-order-QI-construction-2024,carroll2013-journal}, defined to be continuous within a field period, {\em i.e.} $\ell \in [2\pi n/N,2\pi(n+1)/N)$, where $n$ is an integer.  By our convention, field periods begin and end at locations of maximum field strength, which are also points of stellarator symmetry \cite{dewar1998stellarator}.  The signed Frenet frame is the same as the standard Frenet frame except that the normal and binormal vectors, $\n$ and $\b$, are `corrected' by a sign to ensure continuity at locations of minimum field strength where the axis curve can be said to undergo {\it inflection}; see details in Appendix~A of \cite{rodriguez-plunk-second-order-QI-construction-2024} and a more general discussion of such frames in \cite{carroll2013-journal}.  The magnetic axis curves considered here can be described by the following expressions for signed curvature and torsion

\begin{eqnarray}\label{eq:kappa-tau}
    \kappa = \kappa_1 \cos^2(N \ell/2) \sin(N \ell) \sin(N \ell / 2),\\
    \tau = \tau_0 + \tau_1 \cos( N \ell).
\end{eqnarray}
where $\kappa_1$, $\tau_0$ and $\tau_1$ are constants.  The form of $\kappa$ is chosen to fix the order of its zeros at the magnetic field extrema: $2$ at maxima and $3$ at minima.  The signed Frenet frame is then obtained by integrating Eq.~\ref{eqn:FS_eqns}, substituting the above forms of $\kappa$ and $\tau$.  Two of the three free constants are determined by iteratively solving the equations to satisfy the closure conditions.

The number of times the signed normal rotates about the axis is referred to as the \textit{axis helicity} $M$ \cite{rodriguez2022phases,mata2023helicity,rodriguez-plunk-second-order-QI-construction-2024} (also self-linking number \cite{moffatt1992helicity,fuller1999geometric,fuller-writhe}) and we define the per-field-period helicity as $m = M/N$.  Fixing the length of the axis to be equal to $2\pi$, the helicity to be $m = 1/2$, and the field period number $N = 2$, we obtain a family of curves, which we parameterize by $\tau_1$; the other constants $\tau_0$ and $\kappa_1$ are determined by the closure condition.

A special case is obtained in the limit $\tau_1 \rightarrow 0$, where the mean torsion is found to also be zero, $\tau_0 = 0$; note also $\kappa_1 \approx 10.2$ in this case.  This zero-torsion case is the planar figure-8 (a \textit{lemniscate} \cite{lawrence2013catalog}), in particular the following shape: \eightchar.  Because the axis is chosen to begin ($\ell = 0$) along the positive $y$ axis, this curve lies entirely in the $x$-$z$ plane and self-intersects at the origin.

The planar figure-8 curve is obviously not suitable for the magnetic axis of a stellarator due to its self intersection.  The curve must be brought somewhat out of the $x$-$z$ plane to open up space for the plasma volume to form around the magnetic axis (also coils, {\it etc.}), which can be achieved by varying the free parameter $\tau_1$.  To complete our solution to zeroth order in the NAE, the magnetic field strength dependence along the magnetic axis is chosen as

\begin{equation}
  B_0 = 1 + 0.3 \cos(2 \ell) + 0.075 \cos(4 \ell),\label{eq:B0}
\end{equation}
defined such that the mirror ratio is $(B_0|_{\ell = 0}-B_0|_{\ell = \pi/2})/2 = 30\%$ \footnote{Note that the $\ell$-average of $B_0$ here is $1$, making this mirror ratio similar but not exactly equal to the typical definition, $(B_\mathrm{max}-B_\mathrm{min})/(B_\mathrm{max}+B_\mathrm{min}) \approx 0.28$.}, and $B_0^{\prime\prime} = 0$ at field strength minima, {\em e.g.} $\ell=\pi/2$, the latter to promote the formation of a magnetic well; see Section \ref{eduardo-sec}.  With this defined, the conversion between axis length and Boozer toroidal angle $\varphi$ may be made \cite{landreman-sengupta,plunk_landreman_helander_2019}:
\begin{equation}
\frac{d\varphi}{d\ell}= \frac{B_0}{|G_0|},
\hspace{0.3in}
|G_0| = \frac{N}{2\pi}\int_0^{2\pi/N} B_0\,d\ell, \label{eq:l-varphi}
\end{equation}
where $G_0$ is the zeroth-order poloidal current function.

\subsection{Configuration shaping to first order}\label{sec:first-order}

To complete the construction of the configuration to first order, in particular to determine the elliptical shape of the surfaces that form around the magnetic axis, we must solve the so-called `sigma equation' \cite{garren-boozer-1}, depending on several given inputs,

\begin{equation}
    \sigma^{\prime} + (\iotaslash_0 -\alpha^{\prime})\left(\sigma^{2}+ 1 + \bar{e}^2 \right) + 2 G_{0} \tau \bar{e}/B_0 = 0,\label{eq:sigma}
\end{equation}
where $\alpha$ and $\bar{e}$ are functions of $\varphi$ \cite{plunk_landreman_helander_2019,camacho-mata_plunk_2022}, and $\iotaslash_0$ is the rotational transform on axis; note we have assumed zero toroidal current.  The solution to this equation determines the surface shape via the components of the first order coordinate mapping,

\begin{eqnarray}\label{eq:X0_Y0}
    X_{1} &= \sqrt{\frac{2}{B_0}}\sqrt{\bar{e}} \cos{[\theta - \alpha (\varphi) ]}, \\
    Y_{1} &= \sqrt{\frac{2}{B_0}} \frac{1}{\sqrt{\bar{e}}} \left( \sin{[\theta - \alpha (\varphi)]} + \sigma(\varphi) \cos{[\theta - \alpha (\varphi) ]}   \right), \label{eq:Y1}
\end{eqnarray}
which enters the coordinate mapping as follows

\begin{equation}
    \mathbf{x} \approx \mathbf{x}_{0} + \epsilon \left( X_{1}\mathbf{n} + Y_{1}\mathbf{b} \right),
\end{equation}
where $\epsilon=\sqrt{\psi}$.\footnote{This pseudo-radial coordinate differs by a factor of $\sqrt{2/\bar{B}}$ compared to some others \cite{landreman-sengupta,rodriguez-plunk-second-order-QI-construction-2024}.}

Because elongation (of the elliptical cross sections) is a particularly sensitive feature of QI configurations that are derived from NAE theory, we choose to solve the $\sigma$ equation in a non-standard fashion \cite{plunk2024-QI}, using elongation as input, instead of the function $\bar{e}$, which is used conventionally \cite{landreman-sengupta-plunk, plunk_landreman_helander_2019}.  The elongation $E(\varphi)$ of the elliptical cross section is

\begin{equation}
    E(\varphi) = \frac{1}{2}\left(\rho + \sqrt{\rho^2-4}\right),\quad \mathrm{with}\quad \rho = \bar{e} + \frac{1}{\bar{e}}(1 + \sigma^2).
\end{equation}
The ``elongation profile'' is set by the function $\rho$, which for the class of configurations considered here is chosen to have the following form

\begin{equation}
\rho = \rho_0 + \rho_1 \cos(2 \varphi),\label{eq:elongation-input-form}
\end{equation}
where $\rho_0$ and $\rho_1$ are constants.  Note that $\rho=2/\sin 2e$, where $e$ is defined in \cite{rodriguez2023magnetohydrodynamic} as the angle subtended by a right angle triangle with the axes of the ellipse as catheti. Thus, a larger $\rho$ represents a larger elongation of cross sections, and the sign of $\rho_1$ gives growing (positive) or decreasing (negative) elongation away from the point of minimum field strength.  A smooth version of $\alpha(\varphi)$ is used for proper behavior at higher order NAE, as described in \cite{rodriguez-plunk-second-order-QI-construction-2024}.

\section{Defining features of a figure-8 QI stellarator}\label{sec:figure-eight-sequence}

The overall geometry of a figure-8 stellarator can be thought intuitively of as a racetrack shape, bent out of the horizontal plane by inclining each end in opposite senses by an amount that we will call the ``inclination angle'', $\gamma$, following Spitzer \cite{spitzer-1958}.  This angle is zero in the racetrack-limit, and must always be somewhat less than $\pi/2$ (the planar curve limit), to allow for sufficient space in the region where the node of the ``8'' appears. We will use the inclination angle as a simple way to quantify the degree to which a configuration approaches the ideal figure-8 shape, and use $\gamma$ instead of $\tau_1$ to parameterize the class of axis curves introduced in the previous section. 

\begin{figure}
    \centering
    \includegraphics[width=0.6\textwidth]{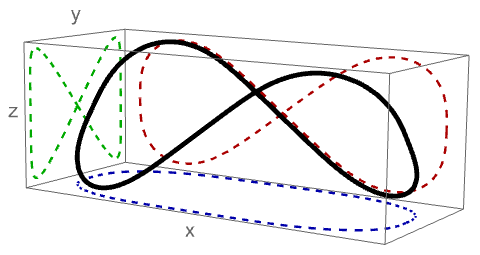}\includegraphics[width=0.3\textwidth]{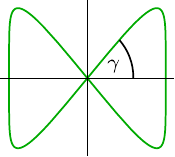}\\
    \vspace{4pt}
    \includegraphics[width=0.9\textwidth]{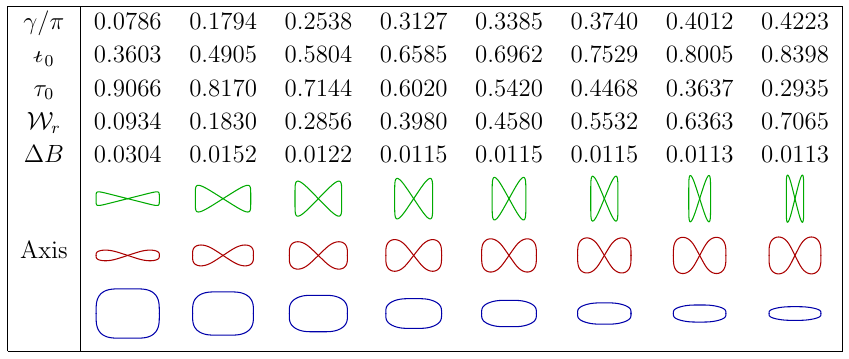}
    \caption{The top two figures show how the inclination angle of the magnetic axis, $\gamma$, is defined using the crossing in the ``bowtie'' ($y$-$z$) projection (green) of the axis curve.  The bottom table describes a sequence of configurations with varying axis inclination, with the final three rows of the table showing different views of the magnetic axis curve for each case: ``bow-tie'' view (green), ``figure-8'' view (red), and ``racetrack'' view (blue).}
    \label{tab:figure8-sequence}
\end{figure}

Varying the inclination angle, we produce eight configurations, all with identical elongation  $\rho(\varphi)$ and on-axis field strength $B_0(\varphi)$.  The elongation (versus $\varphi$) varies from $\sim 6$ at $\varphi = 0$ to $\sim 2.4$ at $\varphi = \pi/2$ (in particular $\rho_0 = 4.5$ and $\rho_1 = 1.7$ in Eqn.~\ref{eq:elongation-input-form}).  A number of other characteristics of these configurations are provided in Figure \ref{tab:figure8-sequence}.  One of those is a proxy for QI error \cite{camacho-mata_plunk_2022}, $\Delta B$, which is observed to monotonically decrease with inclination angle, as observed previously \cite{plunk-simons-2023}.  This quantity is defined as the root-mean-square deviation of $|{\bf B}|$, as obtained from the equilibrium code, from the theoretical value predicted by first order near-axis theory $B_0 + \epsilon B_1$; see Sec.~\ref{sec:well-opt}.

The inclination angle $\gamma$ is a somewhat arbitrary measure, and unsatisfying from a theoretical perspective.  To dig deeper we note a remarkable feature of the class of curves, defined to have total axis helicity $M = 1$.  In the limit as $\gamma$ approaches $\pi/2$, the torsion tends to zero point-wise, implying that the  axis normal manages to undergo a complete rotation without torsion.  In the language of the mathematics of closed space curves (or more precisely `ribbons'), it can be said that such a curve generates its self-linking via {\em writhe} (mostly) instead of {\em twist}, the latter being the integrated torsion \cite{moffatt1992helicity,fuller1999geometric,fuller-writhe}.  The contribution to self-linking from these two sources is expressed by the C\u{a}lug\u{a}reanu-White-Fuller formula \cite{fuller-writhe,moffatt1992helicity},

\begin{equation}
    M = {\cal W}_r + \frac{1}{2\pi}\oint \tau dl,
\end{equation}
whereby the writhe ${\cal W}_r$ is formally defined.  In this case we have defined the ribbon as the magnetic axis curve along with the signed normal vector; see the table in Fig.~\ref{tab:figure8-sequence} for values of ${\cal W}_r$ and average torsion $\tau_0$ for the sequence of axis inclinations.  Intuitively, curves with high writhe are naturally formed by elastic objects like rods or belts, to relieve internal twisting stresses while respecting self-linking, which is topologically constrained.  

This suggests another way to identify a figure-8 stellarator (alternative to using the inclination angle), namely by the feature that the helicity (self-linking) of its magnetic axis largely achieved without torsion, {\it i.e.} it is a curve of large writhe.
This property of the figure-8 has significant consequences for how the stellarator generates rotational transform \cite{pfefferle2018non,mercier-near-axis,rodriguez2022phases}, and for a number of other attractive properties of the configurations, as we will show in what follows.  

According to near-axis formalism, the on-axis rotational transform can be related to helicity and torsion as a solubility condition of Eqn.~\ref{eq:sigma},

\begin{equation}
    \int_0^{2\pi/N} ( \iotaslash_0 - \alpha^\prime)(\sigma^{2}+ 1 + \bar{e}^2) d\varphi = - \int_0^{2\pi/N} 2 \tau G_{0}\bar{e}/B_0  d\varphi.\label{eq:iota-solubility}
\end{equation}
There is both explicit and implicit dependence on $\iotaslash_0$ in this equation, as the function $\alpha$ depends on $\iotaslash_0$.  Indeed omnigenity (assuming stellarator symmetry) implies $\alpha^\prime = \iotaslash_0$, which however cannot generally be satisfied for all $\varphi$ because of the periodicity constraint $\alpha(2\pi) - \alpha(0) = M$ \cite{plunk_landreman_helander_2019, camacho-mata_plunk_2022}.\footnote{This reflects a more general conflict, beyond NAE, between omnigenity and periodicity at field maxima under the assumption of irrational $\iotaslash$ \cite{cary-shasharina}.  In near-axis theory of QI the conflict is addressed by defining $\alpha(\varphi)$ to control the violation of omnigenity \cite{plunk_landreman_helander_2019, camacho-mata_plunk_2022, rodriguez-plunk-second-order-QI-construction-2024}.}  However, in the limit as $\tau \rightarrow 0$, noting the second factor of the integrand on the left-hand-side of Eq.~\ref{eq:iota-solubility} is positive-definite, we do find $\alpha^\prime \approx \iotaslash_0$, and therefore $\iotaslash_0 \approx M$ by periodicity; the trend toward this solution can be observed in Figure~\ref{tab:figure8-sequence}.  Practically speaking, this means that as the figure-8 limit is approached, it becomes possible to satisfy the omnigenity condition more uniformly.  Furthermore, low torsion favors low amplitude of the function $\sigma$ obtained from Eq.~\ref{eq:sigma}, as has been noted before \cite{camacho-mata_plunk_2022}, which controls the tilt of the ellipses \cite{rodriguez2023magnetohydrodynamic}.  Therefore, in the figure-8, the ellipses stay closely aligned with the Frenet frame (see Fig.~\ref{fig:cross-sections-shaping-figure-8-vs-racetrack}), which itself performs very little twist about the axis. 

To summarize, the figure-8 is special in that it generates a relatively large rotational transform with little axis torsion.  Using near-axis theory, a QI figure-8 can be constructed that, as a consequence of low torsion, has exceptionally weak shaping at first order.  As we will find in the following sections, weak shaping also extends, in a particular sense, to higher order.

\section{Susceptibility of the Figure-8 to the formation of a magnetic well}\label{sec:well-opt}

A first issue to investigate for the figure-8 is the question of stability, which was historically one of the main criticisms motivating the switch to classical stellarators.  Stellarator optimization commonly uses proxies for linear stability \cite{Drevlak_2019}.  Here we focus on the most basic one, the so-called magnetic well, $d^2 V/d\psi^2$ \cite{greene1997,freidberg2014ideal} with $V$ the plasma volume, which enters prominently in Mercier's stability criterion for ideal-MHD interchange stability \cite{Mercier_1964}.

\begin{figure}
     \centering
     \includegraphics[width=0.3\textwidth]{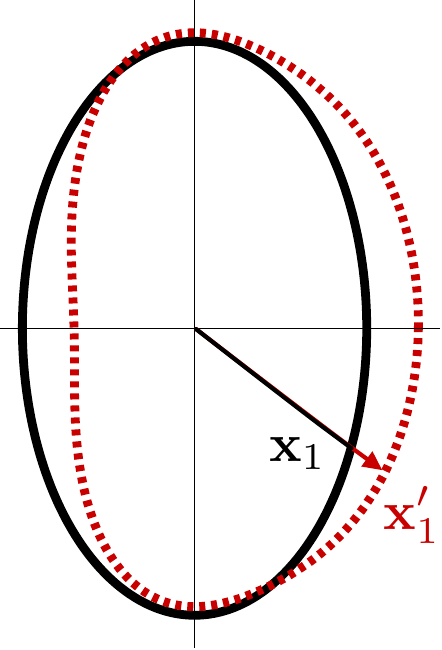}
     \caption{Cartoon of radial deformation, which defines the shape space for the optimization of the magnetic well.  The deformed shape can be written in terms of the original shape (obtained from the first-order NAE) as ${\bf x}_1^\prime = \nu {\bf x}_1$.}
     \label{fig:radial-deformation}
 \end{figure}

To explore the question of stability, we conduct a series of optimizations, based on the set of initial stellarator configurations, constructed with different levels of inclination $\gamma$ as defined in the previous section.  Optimization of the boundary shape is performed using new capabilities of the 3D equilibrium solver GVEC, employing coordinates based on a generalized Frenet frame \cite{Hindenlang_GVEC-Frenet_2022}.  The shape of the boundary is specified as an input by the components of a coordinate mapping, here initially calculated to first order.  We are then able to provide a boundary shape as input to GVEC by modifying the boundary obtained from the NAE, {\em i.e.} ${\bf x}_{\mathrm{mod}} = {\bf x}_0 + \epsilon {\bf x}_1^\prime$ with ${\bf x}_1^\prime = \nu(\theta, \varphi) {\bf x}_1$.  Radial deformation of the first-order shape is thereby achieved by the function $\nu$ whose effect is to move the surface shape along the `radial' direction defined by ${\bf x}_1$, as depicted in Fig.~\ref{fig:radial-deformation}.  The choice $\nu = 1$ corresponds to the unshaped (elliptical) first-order solution; it is the choice that is used to compute the values of $\Delta B$ shown in Figure \ref{tab:figure8-sequence}.  To parameterize $\nu$, a number of uniformly distributed but independent control points are defined in the $\theta$-$\varphi$ plane, taking stellarator symmetry and field-periodicity into account ($\theta$ ranges from $0$ to $2\pi$ and $\varphi$ from $0$ to $\pi$).

The optimization space consists of the value of $\nu$ at the control points, and these values are bounded between $0.9$ and $1.1$ so as to limit the deformation to a maximum of $10 \%$.  The number of control points is chosen ($5$ in the $\theta$ direction and $5$ in the $\varphi$ direction) to control `triangularity' at several values of the toroidal angle, {\em i.e.} the shape freedom that arises at second order in the NAE, which is known to control the magnetic well \cite{Landreman_Jorge_2020, rodriguez2023magnetohydrodynamic}.

A total of eight optimizations are performed, corresponding to the cases presented in Figure \ref{tab:figure8-sequence}.  By limiting the deformation size, global optimization (we use the  Bayesian method in \texttt{Mathematica}) can be used to find the maximum of the target function, which is chosen to be the magnetic well depth (note that positive values indicate stability, opposite to the case for $d^2V/d\psi^2$)

\begin{equation}
w(s) = (V'(0)-V'(s))/V'(0)
\end{equation}
where $s$ is a normalized surface label related to the toroidal flux function, $\psi = s \psi_\mathrm{edge}$.

\begin{figure}
    \centering
    \includegraphics[width=0.47\textwidth]{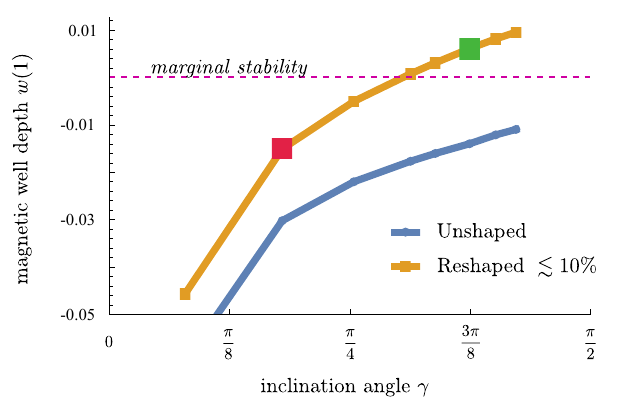}\hspace{0.3cm}\includegraphics[width=0.5\textwidth]{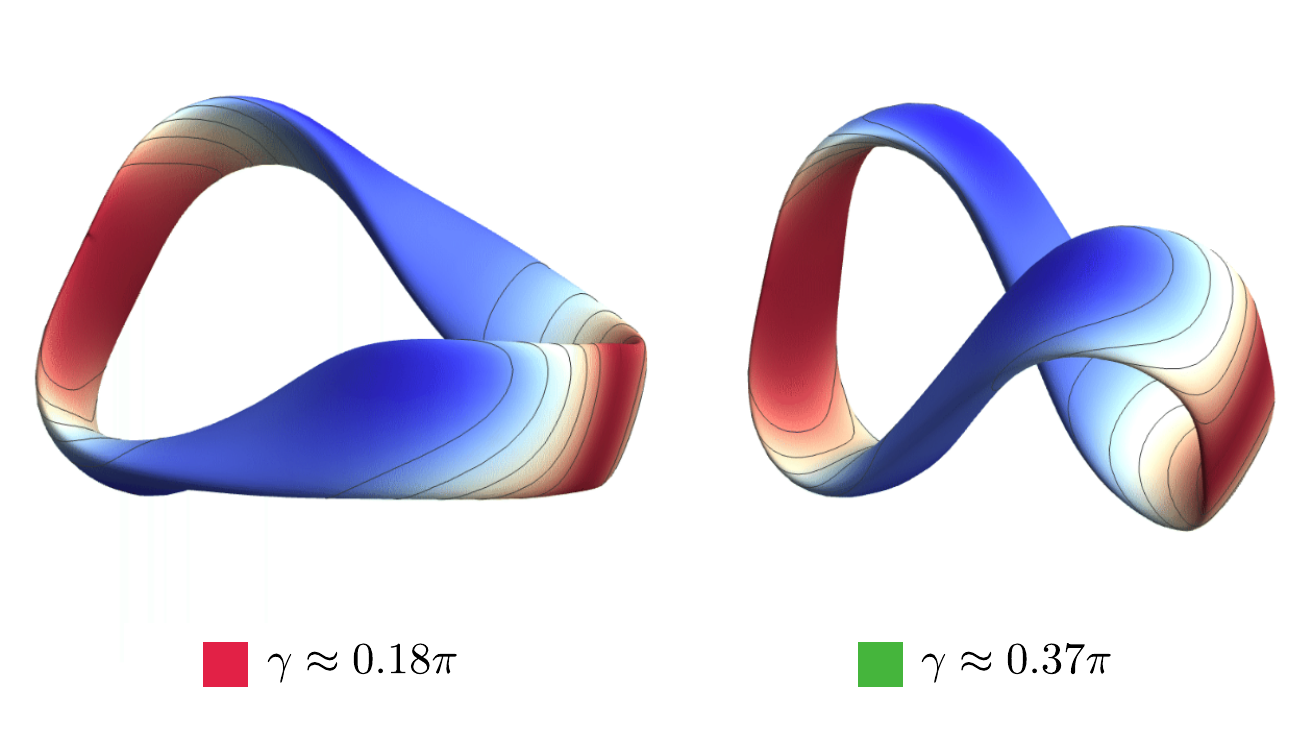}
    \caption{Maximum magnetic well depth versus inclination angle (left).  Visualization of magnetic surface for two cases of low inclination (middle) and high inclination (right).}
    \label{fig:well-vs-inclination}
\end{figure}

\begin{figure}
    \centering
    \includegraphics[height=3.0cm]{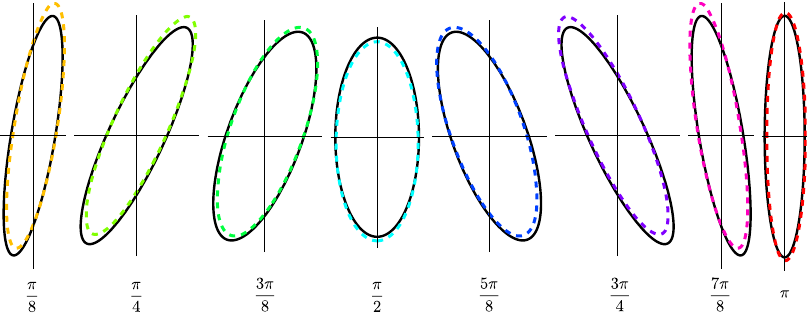}\hspace{1.25 cm}\includegraphics[height=3.0cm]{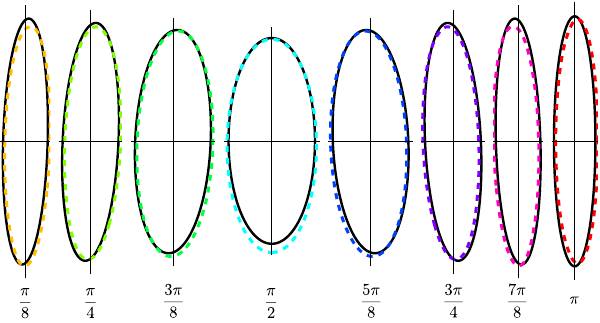}
    \caption{Comparison of magnetic surface shaping for the cases of lowest inclination ($\gamma = 0.0786\pi$; left) and highest inclination ($\gamma = 0.4223\pi$; right).  Cross sections are shown for equally spaced values of $\varphi$, as labeled.  The initial (perfectly elliptical) shapes are shown in black and the final (optimized) shapes are shown in dashed colors.  The shapes are plotted in the Frenet frame, which is why the ellipses tend to align vertically.}
    \label{fig:cross-sections-shaping-figure-8-vs-racetrack}
\end{figure}

The result of these optimization runs is shown in Fig.~\ref{fig:well-vs-inclination}.  It is shown that the maximum achievable well depth is a monotonically increasing function of the inclination angle.  Simultaneously the proxy for QI error $\Delta B$ is observed to monotonically decrease, as shown in Figure \ref{tab:figure8-sequence}.  Note that the final boundary shapes resulting from the optimization are nearly  indistinguishable visually from the initial elliptical shapes, as demonstrated for two extreme values of $\gamma$ in Fig.~\ref{fig:cross-sections-shaping-figure-8-vs-racetrack}

\subsection{Discussion of MHD stability}\label{eduardo-sec}

Practically speaking, the above results show by example that the figure-8 is compatible with a magnetic well, even especially so in some sense.
While these results are encouraging, we would like to probe deeper to understand the reason for the observed tendency, and determine what can be said more generally about the stability of this type of stellarator.

We therefore turn back to near-axis theory, which provides information about the vacuum magnetic well at second order \cite{Landreman_Jorge_2020, rodriguez2023magnetohydrodynamic}, here expressed, assuming vacuum, as $V''=-V'\mathrm{d}\ln\langle B^2\rangle/\mathrm{d}\psi$, recalling that $V'' < 0$ indicates stability, {\em i.e.} outwardly growing magnetic pressure \cite{greene1997,freidberg2014ideal}.

\begin{figure}
    \centering
    \includegraphics[width=\textwidth]{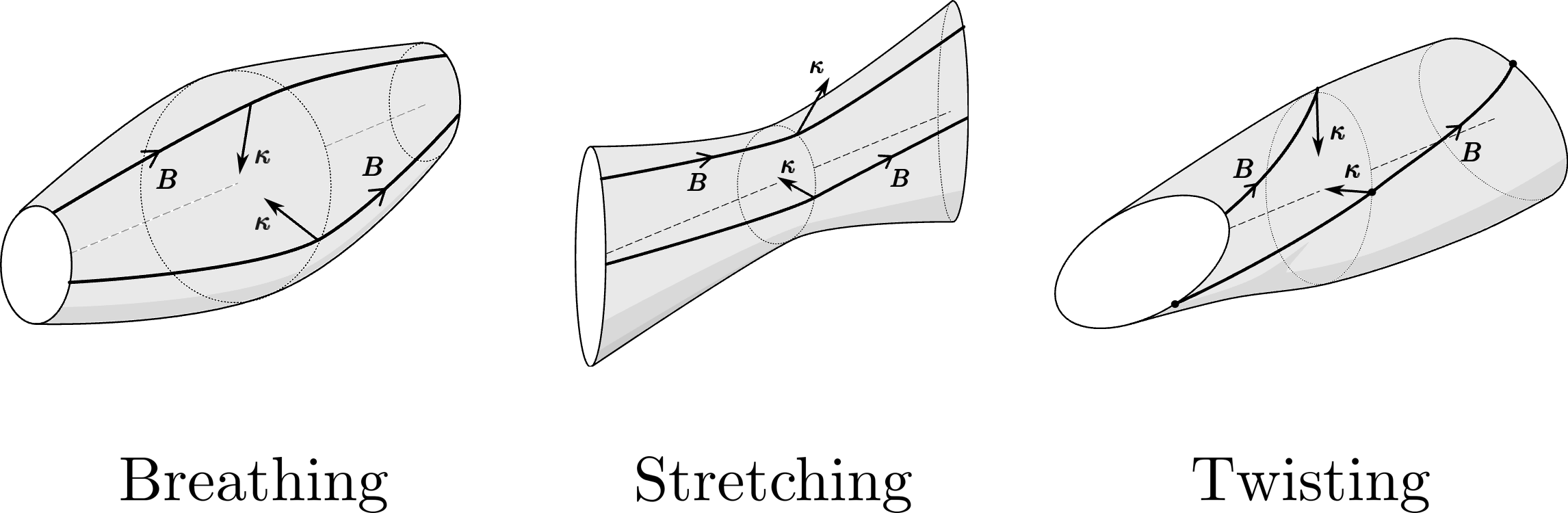}
    \caption{Depiction of three contributions to the magnetic well near locations of field extrema in QI stellarators. The magnetic field strength grows normal to the axis in the direction of the curvature $\pmb{\kappa}$ of the field-lines, and thus the behavior of $\pmb{\kappa}$ can be considered to provide a physical origin to the contributions in Eqn.~\ref{eq:magnetic-well-contributions-QI}.}
    \label{fig:shaping_well_nae}
\end{figure}

Instead of explicitly constructing QI configurations at second order \cite{rodriguez-plunk-second-order-QI-construction-2024}, which, depending on shaping choices, only provides similar anecdotal evidence as already shown above, we turn to more basic considerations.  To see how the properties of the first-order construction ({\em i.e.} the ``unshaped'' configuration) can affect stability, we can examine the locations of zero curvature along the axis, where the field geometry resembles a straight mirror \cite{rodriguez-plunk-second-order-QI-construction-2024, rodriguez2024maximum}.  Here the radial variation of $B$, being constrained by the vacuum expression $\nabla_\perp(B^2/2) = B^2\pmb{\kappa}$, with $\pmb{\kappa} = (\pmb{B}/B)\cdot\pmb{\nabla}(\pmb{B}/B)$, is completely determined by first-order properties, and can be expressed to leading order near the magnetic axis as

\begin{equation}
    \frac{1}{2B}\left\langle\frac{dB}{d\psi}\right\rangle_\theta = \frac{\rho}{4}\left(\frac{d\varphi}{dl}\right)^2 \left [-\frac{B_0^{\prime\prime}}{2B_0} + \frac{1}{2} \frac{\rho^{\prime\prime}}{\rho} -\left(\tau \frac{dl}{d\varphi}\right)^2 \right],\label{eq:magnetic-well-contributions-QI}
\end{equation}
where primes denote derivatives with respect to $\varphi$, the angled brackets an average over $\theta$, and the other symbols have been defined in Sec.~\ref{sec:nae_construction}.  This expression comes from Eqn.~B4 of \cite{rodriguez2024maximum}, assuming vacuum conditions, using $\sigma = 0$ at locations of stellarator symmetry, and re-expressed in terms of elongation (more precisely $\rho$) to be consistent with the approach to first-order construction used here; see Sec.~\ref{sec:first-order}.  The three resulting terms within the bracket (note the signs) can be interpreted in the way depicted in Fig.~\ref{fig:shaping_well_nae}.\footnote{This behavior is purely based on equilibrium considerations.  The picture changes if, in addition to this, the field is required to satisfy omnigenity beyond first order, as discussed in \cite{rodriguez2024maximum}.} These diagrams show three fundamental ways in which the geometry of the surface, through the curvature vector $\pmb{\kappa}$ of field lines, affect the $\theta$-average of $dB/d\psi$.
\par
\par
``Breathing'' (first term) is the flux-preserving variation in cross-sectional areas due to variation of $B_0$, which makes a destabilizing contribution at the field minimum, and stabilizing at the field maximum. To minimize its detrimental effect, $B_0$ (Eqn.~\ref{eq:B0}) was chosen to be flat in the neighborhood of the minimum. 

``Stretching'' (second term) is the variation of elongation, measured here by $\rho(\varphi)$, which on average favors stability when the elongation grows away from the inflection point, {\em i.e.} for $\rho_1>0$ as chosen in Sec.~\ref{sec:first-order}. This positive contribution is identical throughout the sequence of configurations in Section \ref{sec:figure-eight-sequence}.  

The remaining effect,``Twisting'' (third term), is the destabilizing effect that axis torsion has via ellipse rotation.  As discussed in Section \ref{sec:figure-eight-sequence}, the reduction of twist, in exchange for increased writhe, is the key geometric feature distinguishing the magnetic axis of the figure-8, and here we find it is the key feature underlying the observed tendencies toward stability.  In short, low level of twisting from axis torsion makes the figure-8 intrinsically more stable at field extrema, so that a relatively small amount of additional shaping elsewhere is needed to produce a global magnetic well.

\section{A Quasi-Isodynamic Figure-8 Stellarator with Planar Coils}\label{sec:QI-figure-eight-planar-coils}

Historically speaking a second advantage (besides stability) that the classical stellarator was thought to have over the figure-8 was the relative simplicity of construction \cite{bishop-project-sherwood}.  To address this issue in the context of modern stellarators, a key question is that of coil complexity.  As a case study we therefore consider a single figure-8 configuration on which to perform coil optimization.  This configuration, distinct from the ones of the previous section, has axis inclination $\gamma = 0.374 \pi$, elongation ranging from approximately $6.65$ at the field maximum, to $1.56$ at the field minimum, $(\rho_0,\rho_1) = (4.5, 2.3)$, and a rotational transform at first order of $\iotaslash_0 \approx 0.7$.

\begin{figure}
    \centering
    \includegraphics[width=0.9\textwidth]{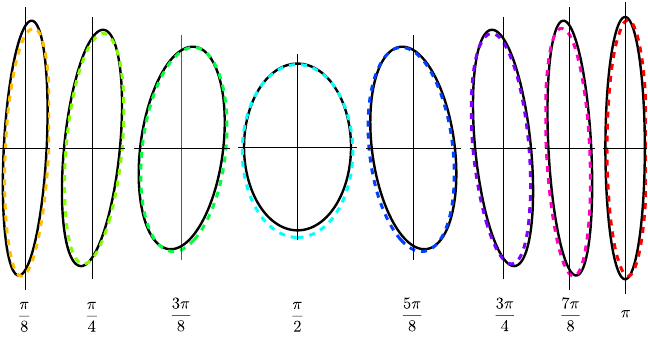}
    \caption{Cross sections are shown for equally spaced values of $\varphi$, as labeled.  The initial (perfectly elliptical) shapes are shown in black and the final (optimized) shapes are shown in dashed colors.  The shapes are plotted in the Frenet frame, which is why the ellipses are mostly aligned vertically.}
    \label{fig:cross-sections-shaping}
\end{figure}

The constructed boundary shape was re-optimized for a magnetic well as described in the previous section, resulting in a well depth $w(1) = 0.8\%$.  To show how the surfaces actually were deformed to achieve this, a set of cross sections at uniformly spaced values of $\varphi$ are shown in Fig.~\ref{fig:cross-sections-shaping}.  The simplicity of these cross sections, showing no visible hint of the expected ``bean'' shape, seem to have strong consequences on the coil set that is obtained during optimization, as presented below.

\begin{figure}
    \centering
    \includegraphics[width=0.4\textwidth]{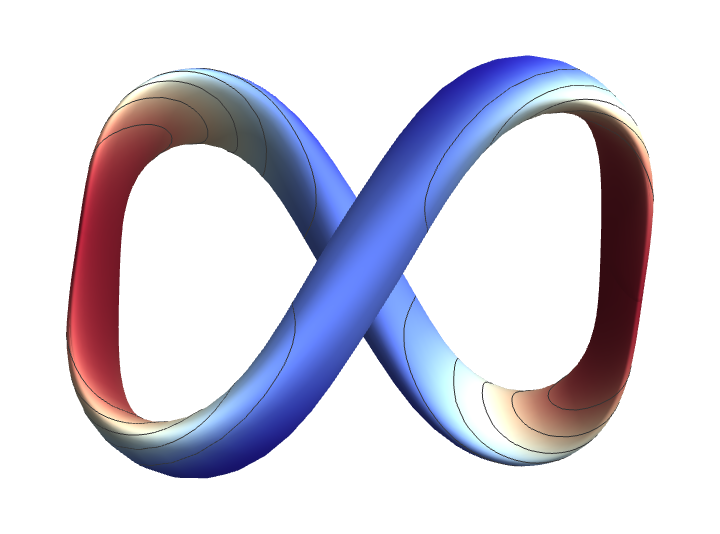}\includegraphics[width=0.24\textwidth]{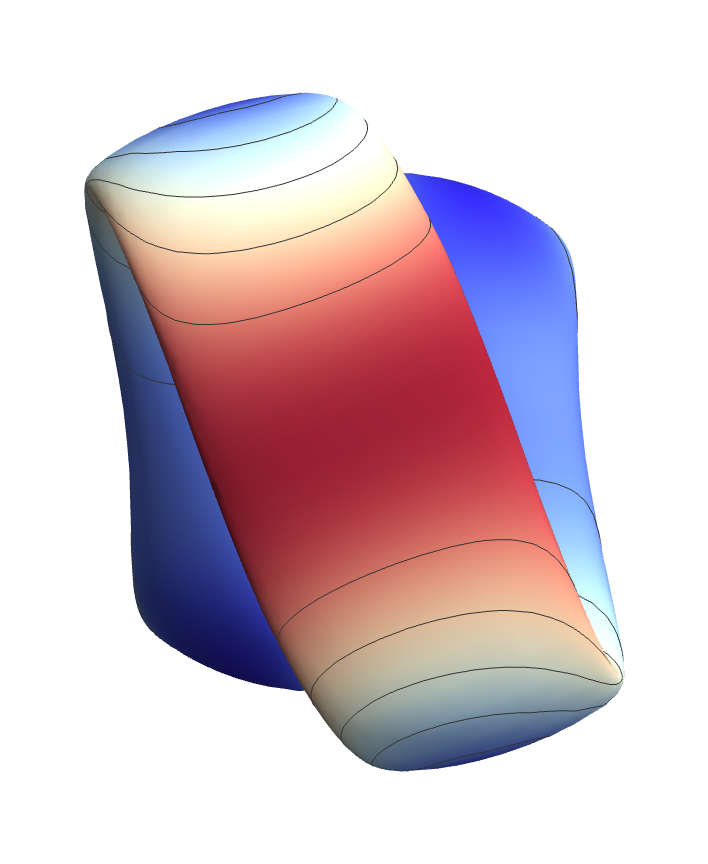}\includegraphics[width=0.4\textwidth]{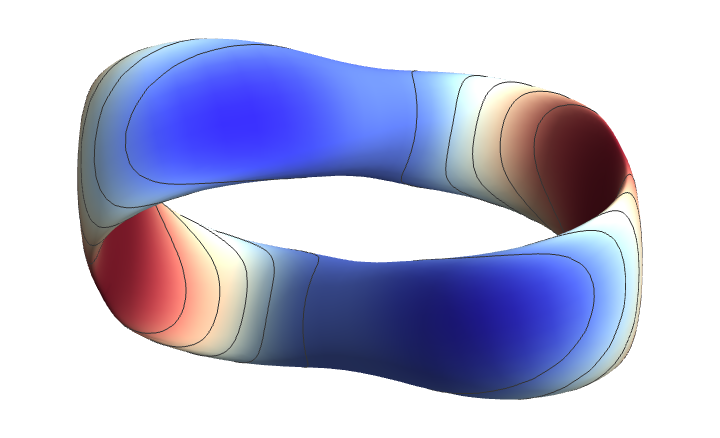}
    \caption{A quasi-isodynamic figure-8 stellarator from three viewpoints: figure-8 view, bow-tie view, and racetrack view.  The aspect ratio is set to $8$, and the magnetic well depth is $0.8\%$.  The colors indicate field strength, from high (red) to low (blue), according to the first-order NAE construction.}
    \label{fig:QI-Fig-8}
\end{figure}

A view of the full configuration is shown in Fig.~\ref{fig:QI-Fig-8}; the configuration bears a striking resemblance to the early conceptual ``Model C'' design, which was never built \cite{Model-C-Report-PMS-19}. Note that the optimization was performed at aspect ratio $8$ but it was decided afterward to increase the aspect ratio of the resulting configuration to $10$ in order to allow more space to fit the coils.

\subsection{Coil optimization}

Coil designs were carried out using the ONSET \cite{onset} suite of codes. Because a first analysis using NESCOIL \cite{nescoil,nescoil_comp} had indicated that the coils would deviate only slightly from a planar shape, the optimization was carried 
out using planar coils only.  Two types of coils were used: a small coil type with reduced current close to the node of the ``8'', designed in size so as to avoid intersection of opposing coils, and a larger coil type for the remainder of the configuration so as to minimize the modular ripple.  The small coils are somewhat closer to the plasma boundary and set the minimum coil-to-boundary distance, which is about half the average minor radius of the configuration.

\begin{figure}
    \centering
    \rotatebox{90}{\includegraphics[width=0.75\linewidth]{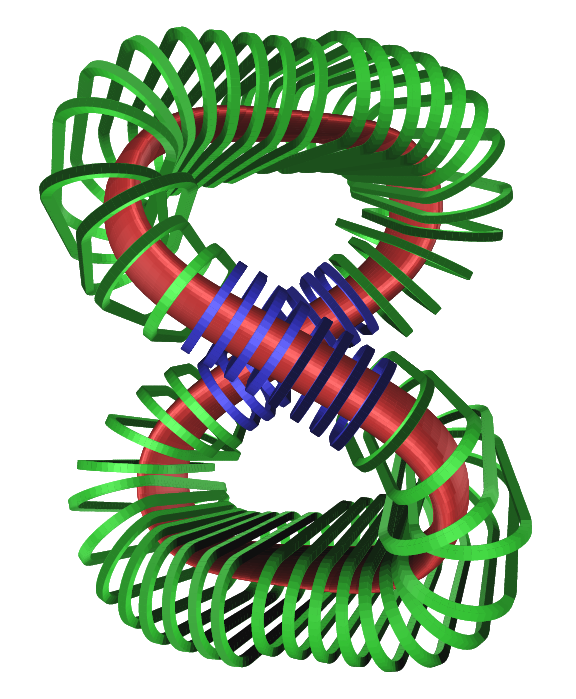}}
    \caption{Racetrack coil set.  There are only two different coil shapes, corresponding to small (blue) and large (green) sizes, and a total of 64 coils in the design, of which only a quarter (16) can be considered independent (the number present in one half field-period).}
    \label{fig:coilset_tilt}
\end{figure}

\begin{figure}
    \centering
    \includegraphics[width=0.41\linewidth]{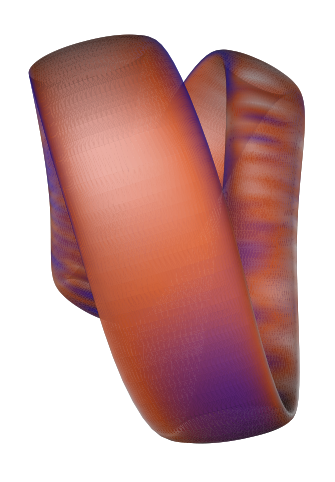}
    \includegraphics[width=0.44\linewidth]{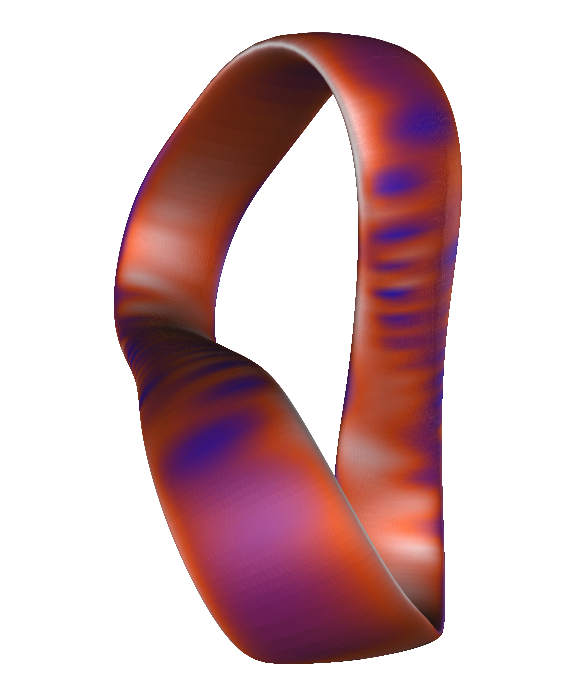}
    \caption{Magnitude of the field error (${\bf Bn}$) on the design plasma boundary as viewed from two different perspectives.  The maximum value is 5.1\% (purple) and the mean is 1.4\%. The colormap encodes max ($B_n$) as blue and min ($B_n$) as white.}
    
    \label{fig:field_error}
\end{figure}

An analysis of the field error (Fig.~\ref{fig:field_error}) shows that the modular field error gradually stops being the dominant cause of field error, starting at a design point with 16 independent coils. On the right hand side of Fig.~\ref{fig:field_error} poloidally elongated structures are identified on the long straight sections that clearly are the effect of the modular ripple. On the left hand side, however, wide areas of normal magnetic field error are seen on the outboard side of the short straight sections, similar in magnitude to the ripple seen in the long sections.

The coil set presented here achieves a maximum field error of 5.1\% and a mean error if 1.4\%. These values tend to result in a good alignment of the coil-generated flux surfaces with those of the design configuration.

\begin{figure}
    \centering
    \includegraphics[trim=0 0 100 350,clip,width=0.6\linewidth]{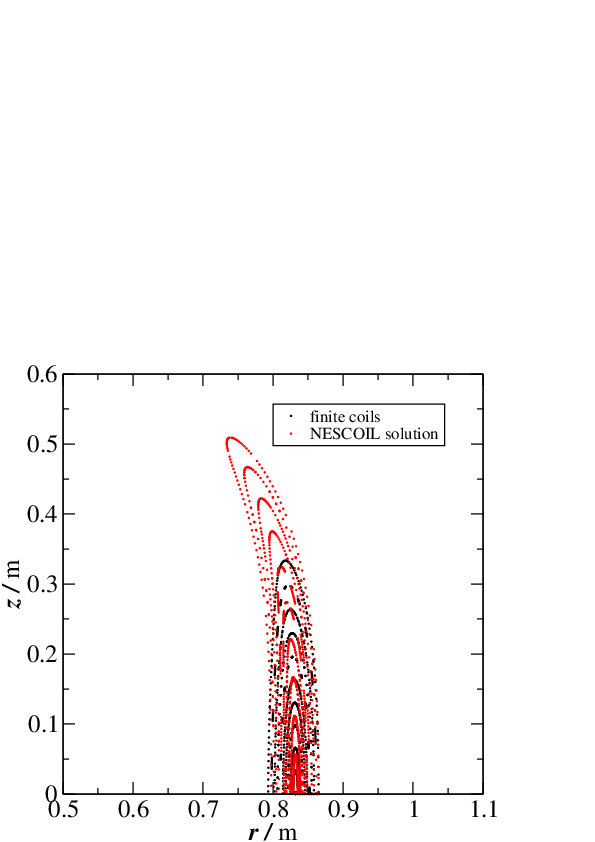}    \caption{Poincar\'{e} plots of the design field (red) and of the coils field (black).  Note that the `bean' shape of the surfaces in the NESCOIL solution arises because the cross sections are made in the cylindrical $R$-$z$ plane.  This can be compared with the almost perfectly elliptical (and less elongated) shapes obtained when cutting perpendicular to the magnetic axis, as shown in Figure \ref{fig:cross-sections-shaping}.}
    \label{fig:poinc}
\end{figure}

Fig. \ref{fig:poinc} finally compares the Poincar\'{e}
data for the design field and for the field from the coil set. The full design field was obtained from an accurate NESCOIL run using the design plasma boundary. The elongation of the flux surfaces obtained from the coils field is lower than that obtained with the accurate NESCOIL solution.
It is expected that this will improve as 
coil sets with lower field error are found.

\section{Conclusion}

In this paper, we have revisited the idea of Spitzer's figure-8 stellarator, re-imagining it as a modern quasi-isodynamic design.  Historically, the brief investigation of the figure-8 concept brought forth concerns about plasma stability, and the alternative that emerged, the classical stellarator, was also regarded as simpler to build.  The initial results found here, including high susceptibility to a stabilizing vacuum magnetic well, and a planar set of coils (the first QI design achieving this), hint that a QI figure-8 stellarator might actually have strengths with respect to both stability and ease of construction.

A key feature linking both properties is the weak shaping of the plasma boundary, which we have traced to a fundamental geometric property at the heart of figure-8 geometry: the magnetic axis curve at the core of the device manages to rotate without twisting.  This means that its mostly elliptical shape stays closely aligned with the smoothly moving frame of the axis, and the lack of twist means that very little high-order shaping ({\em e.g.} triangularity or `bean' shaping) is needed to produce a stabilizing magnetic well.

Intuitively, weak shaping appears to have dramatic consequences on the coil design, with the case investigated here requiring only planar `racetrack' shapes.  The arguments of the paper ultimately connect this design feature with low axis torsion, which may relate to previous observations showing the link between axis geometry and coil complexity \cite{HUDSON_2018}.  Future work could build on this example, or lead to general strategies to reduce coil complexity, with this being one of the most frequently cited criticism of optimized stellarators.

The recent extension of theoretical methods, chiefly near-axis theory, has lead researchers toward unusual shapes like the figure-8 that might otherwise have been neglected.  Until now, numerical tools such as equilibrium codes like VMEC \cite{Hirshman}, which have formed the backbone of numerical stellarator optimization, were not well-suited for many such shapes.  It is ironic that these tools now require updating to investigate the very first, and in some sense simplest stellarator concept.  The extension of the GVEC code \cite{Hindenlang_GVEC-Frenet_2022} is a first step in this direction.  The `re-optimization' near-axis configurations for a magnetic well is only a first demonstration of this kind of application.  The figure-8 is a particularly pathological case for using cylindrical coordinates, and the benefits of doing optimization may extend to more conventional cases, providing a condensed (lower-dimensional) representation for shapes that have posed a challenge for optimization.

The final design presented here was chosen to have an aspect ratio of $10$, for direct comparison with typical for QI stellarators, {\em i.e.} to show the differences that arise at a similar design point.  The inclination of the axis (``figure-8-ness'') was only limited to allow space for the planar coils at the location of minimum field strength.  Practically speaking, a future figure-8 device could be explored for which such parameter choices are adjusted, for instance if more space for the coils is desired, or compactness (smaller aspect ratio) is desired for improved confinement, with some trade-offs in terms of coil distance and/or complexity.

Moving forward there are many practical questions to tackle in all the usual areas, like (global MHD) stability, turbulence properties, confinement, plasma currents, and resilience of the equilibrium to finite plasma pressure.  There are also some fundamental theoretical avenues to investigate, such as a broader search for configurations with low integrated torsion or points of zero torsion that may show some of the benefits observed here.

Overall, the results of this paper seem to cast the figure-8 in a new light.  With more work a fusion-relevant design might be found, or the lessons learned might inspire a related design that borrows some of the strengths of the figure-8.

\section*{Acknowledgments}

We thank Per Helander and Carolin N{\"u}hrenberg for helpful discussions.  E. R. was supported by a grant of the Alexander-von-Humboldt-Stiftung, Bonn, Germany, through a postdoctoral research fellowship.


\section*{References}
\bibliographystyle{vancouver}
\bibliography{neo-spitzer}{}

\end{document}